\documentclass[journal]{IEEEtran}
\usepackage[dvipdfmx]{graphicx,xcolor}
\usepackage{epsfig}

\usepackage[T1]{fontenc}
\usepackage{lmodern}
\usepackage{textcomp}
\usepackage{latexsym}
\usepackage{cite}
\usepackage{bm}
\usepackage{amsmath}
\usepackage{amsfonts}

\usepackage[subrefformat=parens]{subcaption}
\usepackage{multirow}
\usepackage{booktabs}
\usepackage{array}

\newcommand{\vct}[1]{\mbox{\boldmath{$#1$}}}

\newcommand{\ee}        {\mathrm{e}}
\newcommand{\jj}        {\mathrm{j}}
\newcommand{\dd}        {\mathrm{d}}
\newcommand{\TT}        {\mathrm{T}}

\begin{document}
\title{Radar-Based Estimation of Human Body Orientation Using Respiratory Features and Hierarchical Regression Model}

\author{Wenxu~Sun, Shunsuke~Iwata,Yuji~Tanaka,~\IEEEmembership{Member, IEEE}, 
 and Takuya~Sakamoto,~\IEEEmembership{Senior Member, IEEE}
  \thanks{W.~Sun, S.~Iwata, Y.~Tanaka, and T.~Sakamoto are with the Department of Electrical Engineering, Graduate School of Engineering, Kyoto University, Kyoto 615-8510, Japan.}
}
\markboth{}%
{Sun \emph{et al.}: Radar-Based Estimation of Human Body Orientation Using Respiratory Features and Hierarchical Regression Model}

\maketitle
\begin{abstract}
  This study proposes an accurate method to estimate human body orientation using a millimeter-wave radar system. Body displacement is measured from the phase of the radar echo, which is analyzed to obtain features associated with the fundamental and higher-order harmonic components of the quasi-periodic respiratory motion. These features are used in body-orientation estimation invoking a novel hierarchical regression model in which a logistic regression model is adopted in the first step to determine whether the target person is facing forwards or backwards; a pair of ridge regression models are employed in the second step to estimate body-orientation angle. To evaluate the performance of the proposed method, respiratory motions of five participants were recorded using three millimeter-wave radar systems; cross-validation was also performed. The average error in estimating body orientation angle was 38.3$^\circ$ and 23.1$^\circ$ using respectively a conventional method with only the fundamental frequency component and our proposed method, indicating an improvement in accuracy by factor 1.7 when using the proposed method. In addition, the coefficient of correlation between the actual and estimated body-orientation angles using the conventional and proposed methods are 0.74 and 0.91, respectively. These results show that by combining the characteristic features of the fundamental and higher-order harmonics from the respiratory motion, the proposed method offers better accuracy.
\end{abstract}


\IEEEpeerreviewmaketitle

\section{Introduction}
\IEEEPARstart{T}{he} importance of constant monitoring of vital signs including respiration has been globally recognized because of the recent respiratory infection pandemic as well as increasingly high demand for monitoring babies, infants, and senior citizens \cite{bib:Ali2021, bib:Massaroni2021, bib:Singh2020}. Among various sensors for monitoring respiration, radar-based sensors have an advantage when constant monitoring is required because the discomfort when wearing a sensor is no longer present \cite{bib:Muragaki2017, bib:Rivera2006, bib:Su2019}. In addition, a single radar system suffices in monitoring multiple people simultaneously, a convenience that cannot be achieved by conventional contact-type sensors \cite{bib:Lu2019, bib:Novak2016, bib:Li2012, bib:Koda2021}.

Radar-based respiratory measurements use the phase of the radar echo signal to estimate body displacement produced through respiratory motion. The waveform of these displacements depends on body orientation, which affects the accuracy in determining the characteristic features of respiratory motion \cite{bib:Li2006, bib:Noguchi2013, bib:Wang2016, bib:Chen2019}. Noguchi \textit{et al}. \cite{bib:Noguchi2013} reports the difficulty in estimating respiration rate accurately when the target person is facing away from the radar antennas. Moreover, body orientation also affects the accuracy of radar-based determination of heartbeat rates \cite{bib:Wang2013,bib:Sakamoto2021}, indicating that an accurate estimation of body orientation leads to better accuracy in radar-based measurements of respiration and heartbeat.

Of the existing studies on radar-based estimation of the human-body orientation \cite{bib:Li2014, bib:Yang2020}, Li \textit{et al.} \cite{bib:Li2014} proposed a method based on measurements of micro-Doppler effects produced by movement of the arms during specific actions; Yang \textit{et al}. \cite{bib:Yang2020} proposed an alternative method based on the random forest algorithm to estimate body orientation. These methods, however, can only estimate approximate body orientations from six or eight directions.

In this study, we propose a radar-based accurate method to estimate body orientation using a hierarchical regression model with respiratory features calculated from body-displacement components in the frequency domain. Unlike conventional methods, our proposed method uses respiratory waveforms instead of simply amplitudes. The proposed method is also expected to contribute to the establishment of a mathematical model that describes the relationship between respiratory displacement and human-body orientation. We evaluated the effectiveness of the proposed method quantitatively using radar data obtained in a study involving five participants.

\section{Radar-Based Respiratory Measurement and Proposed Method}
\subsection{Respiratory Measurement Using Radar}
We used a radar system with a multiple-input and multiple-output (MIMO) array with three transmitting and four receiving elements spaced respectively $2\lambda$ and $\lambda/2$ apart, where $\lambda$ is the wavelength. This MIMO array can be approximated as an $N$-element virtual array with $N=12$ and element spacing of $\lambda/2$. Let $s_n(t,\tau)$ be the signal received by the $n$-th virtual element, where $t$ is the slow time and $\tau$ is the fast time; a complex-valued radar image $I'(r,\varphi,t)$ is obtained as $I'(r,\varphi,t)=\sum_{n=1}^{N} w_n(\varphi) s_n(t,2r/c)$, with $r$ denoting range and $\varphi$ the azimuth counter-clockwise angle with respect to the array's normal direction. Here, $w_n(\varphi)=\alpha_n\ee^{\jj \pi (n-1) \sin\varphi}$ is a beamforming weight with Taylor coefficient $\alpha_n$. The polar coordinates $(r,\varphi)$ when converted to Cartesian coordinates $(x,y)=(-r\sin\varphi,r\cos\varphi)$ give the complex-valued radar image $I_0(\vct{r},t)$ at position $\vct{r}=(x,y)$, where the $x$-axis is in the direction of the array baseline. 

The complex radar image $I_0(\bm{r},t)$ contains desired echoes from targets and undesired static clutter reflected from stationary objects. Removing the static clutter yields a clutter-free radar image $I(\bm{r},t)$ obtained by subtracting the average over a time duration $T$ as $I(\bm{r},t)=I_0(\bm{r},t)-({1}/{T})\int_{t-T}^t I_0(\bm{r},\tau)\mathrm{d}\tau$. Target position $\bm{r}_0={\arg\max}_{\vct{r}} \int_0^T |I(\vct{r},t)|^2\dd t$ is estimated. The body displacement $d(t)$ is then obtained as $d(t) = ({\lambda}/{4\pi})\mathrm{unwrap}(\angle I(\bm{r}_0,t))$, where $\angle(\cdot)$ signifies the phase of a complex number, and $\mathrm{unwrap}(\cdot)$ the phase unwrapping operator. In the next section, we propose a novel method for estimating the body orientation using $d(t)$.

\subsection{Proposed Method for Body Orientation Estimation}
The fundamental frequency $f_0$ of the respiration can be estimated from the Fourier transform $D(f)=\mathcal{F}[d(t)]$, where $\mathcal{F}[\cdot]$ denotes the Fourier transform operator. Using $D(f)$ and $f_0$, we define a feature vector $\vct{x}=[1, x_1,x_2,x_3,x_4,x_5]^\TT$, where $x_1=|D(f_0)|$ is the amplitude of the fundamental, $x_2=|D(2f_0)/D(f_0)|$ the amplitude of the second harmonic, $x_3=\angle [D(2f_0)/D(f_0)]$ the phase of the second harmonic, $x_4=|D(3f_0)/D(f_0)|$ the amplitude of the third harmonic, and $x_5=\angle [D(3f_0)/D(f_0)]$ the phase of the third harmonic. Note that the second and third harmonics are normalized by the fundamental frequency component $D(f_0)$.

Note that in many cases, human respiratory movements are bilaterally symmetrical, and thus, distinguishing the sign of $\theta$ is fundamentally impossible (i.e., the respiratory motion is almost the same for $\theta$ and $-\theta$). Therefore, we impose a condition $0^\circ \leq \theta \leq 180^\circ$, where $\theta = 0^\circ$ and $180^\circ$ corresponding to a body orientation facing forward and backward respectively with respect to the radar antenna.

One of the simplest methods to estimate body-orientation angle is the linear regression model; the estimate $\hat{\theta}$ is calculated using $\hat{\theta}=\vct{w}^\TT \vct{x}$, where $\vct{w}$ is a weight vector (a superscript $\mathrm{T}$ signifies the transpose operator), which is deemed to be a one-step method in the following sections. Instead of this one-step method with a simple linear regression model, our proposed method is based on a hierarchical regression model that comprises two steps. The first step determines whether $\theta$ belongs to class 1 ($0^\circ \leq \theta< 90^\circ$) or class 2 ($90^\circ \leq \theta \leq 180^\circ$); the second step estimates the body-orientation angle $\hat{\theta}$. Because respiratory motions seen from a front-facing and a back-facing body are significantly different \cite{bib:Groote1997}, the first step becomes a binary classification for which we use a logistic regression model and adopt the logit link function $\mathrm{logit}(z) = \log \left[z/(1-z)\right]$ for $0<z<1$. This function is used to construct a generalized linear regression model. The output class of the classifier is determined based on the sign of the log odds, which is defined as $\mathrm{logit} \left( p \right) = \vct{\beta}^\TT\vct{x}$, where $p$ is the probability of the feature vector belonging to class 1, $\vct{\beta}=[\beta_0, \beta_1,\beta_2,\cdots,\beta_M]^\TT$ is a weight vector, for which we set the dimension of the feature vector to $M=5$ except for a constant term.

In the second step, we construct two ridge-regression model and estimate $\hat{\theta}$ using $\hat{\theta}=\vct{w}_1^\TT \vct{x}$ (if $\mathrm{logit} \left( p \right)\geq 0$) or $\hat{\theta}=\vct{w}_2^\TT\vct{x}$ (if $\mathrm{logit} \left( p \right)<0$), where one of the two models is used depending on the output class (class 1 or class 2) of the first step. To determine the weight $\vct{w}=\vct{w}_1$ or $\vct{w}_2$, we solve the optimization problem:
\begin{equation}
\min_{\vct{w}} \sum_{\ell=1}^{L} \left|\theta_\ell - \vct{w}^\TT \vct{x}_\ell\right|^2
+ \gamma \left|\vct{w}\right|^2,
\label{eq:step2}
\end{equation}
where $\vct{x}_\ell$ $(\ell = 1, \cdots, L)$ denotes the feature vector corresponding to the $\ell$-th body orientation angle $\theta_\ell$, $\gamma$ a regularization parameter set to $\gamma=0.05$ based on empirical results.

\section{Experimental Performance Evaluation of the Proposed Method}
\subsection{Experimental Setup}
\begin{figure}[tb]
    \centering
    \includegraphics[width =0.6\linewidth]{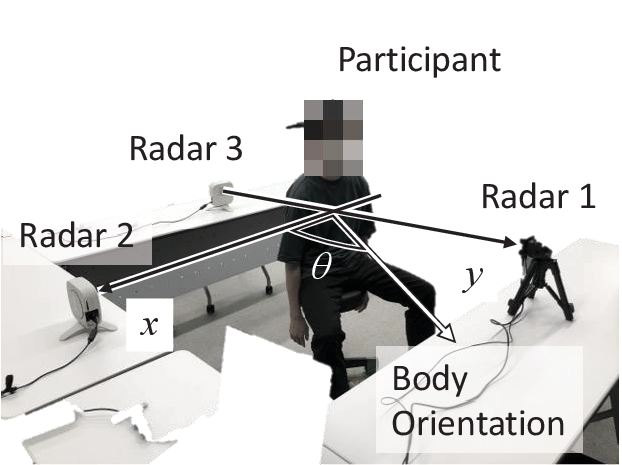}
    \caption{Experimental setup showing the locations of the three radar systems around the participant. The body orientation angle $\theta$ is defined as the angle between the $x$-axis vector and the normal vector directed away from the front of the torso.}
    \label{Fig:Experimental_setup.eps}
\end{figure}

We used a frequency-modulated continuous-wave radar system with a center frequency of 79 GHz, a center wavelength $\lambda=3.8$ mm, an occupied bandwidth of 3.9 GHz, a range resolution of 44 mm, and a slow-time sampling frequency of 10 Hz. The beam-width of the radar array elements are $\pm4^\circ$ and $\pm35^\circ$ in the E- and H-planes, respectively. Several respiratory displacement waveforms obtained from five male participants, whose ages ranged between $21$--$23$ years, were recorded from different angles; see Fig.~\ref{Fig:Experimental_setup.eps}. The protocol involved seating the participant and instructing them to breathe naturally. Three radar systems were positioned approximately 1.0 m away from the participant. Each recording of the waveform lasted 40 s, and after each recording, the participant's body orientation angle $\theta$ was changed sequentially by $10^\circ$ (Fig.~\ref{Fig:Experimental_setup.eps}). For each participant, a total of 57 waveforms from the three radar systems were recorded.

\begin{figure}[tb]
  \begin{minipage}[b]{0.49\linewidth}
    \centering
    \includegraphics[width = 1\linewidth]{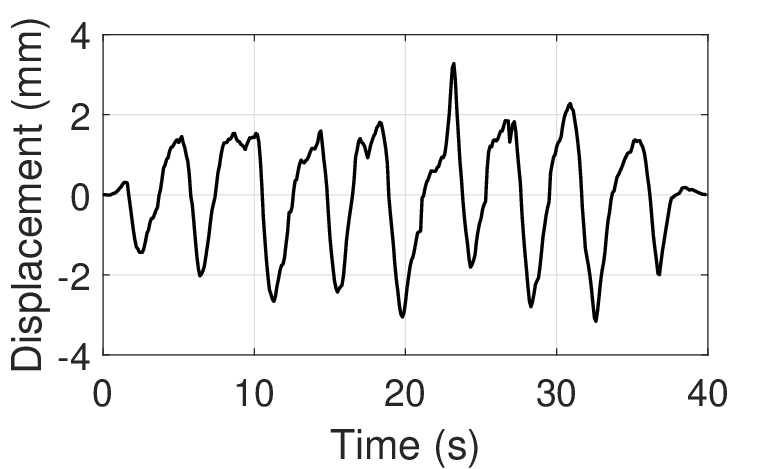}
  \subcaption{}
  \end{minipage}
  \begin{minipage}[b]{0.49\linewidth}
    \centering
    \includegraphics[width = 1\linewidth]{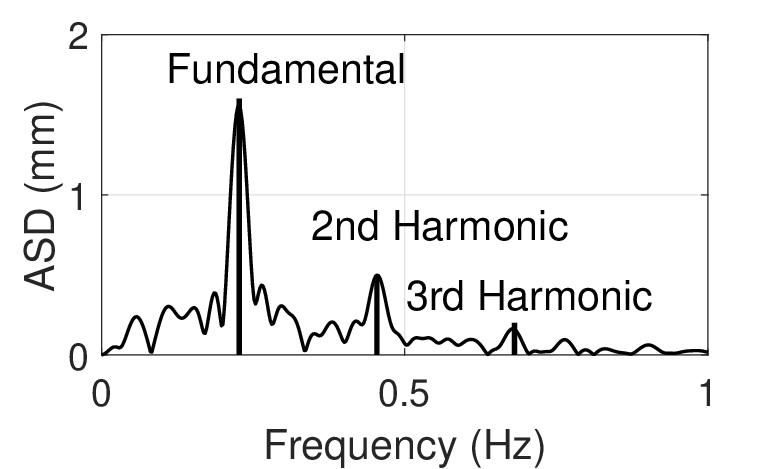}
  \subcaption{}
  \end{minipage}
  \caption{Typical respiratory displacement waveform (a) and its amplitude spectral density (ASD) function (b).}
  \label{Fig:Waveform.eps}
\end{figure}

The feature vector $\vct{x}$ was extracted from the amplitude spectral density (ASD) function for each of the $L=19$ body orientations. Fig.~\ref{Fig:Waveform.eps} shows an example of a respiratory waveform and its ASD function. Note that the phase of the fundamental component $\angle D(f_0)$ was not used as a feature, because the respiratory phase cannot be controlled in the measurement. In the first step, we evaluated the effectiveness of the binary classifier $\mathrm{logit} \left( p \right) = \vct{\beta}^\TT\vct{x}$ using a chi-squared test and a receiver operation characteristic (ROC) analysis for various combinations of features $\vct{x}$. We also evaluated the performance of a simple method in which only the amplitude of the fundamental frequency component $x_1 = |D(f_0)|$ is used; its performance is discussed later. To evaluate the accuracy of the various methods, the coefficient of correlation (CC) $\rho=\langle (\hat{\theta}-\langle \hat{\theta} \rangle)(\theta - \langle \theta\rangle) \rangle$ between $\theta$ and $\hat{\theta}$ (i.e., the actual and estimated body orientation angles) is used; here $\langle\cdot\rangle$ denotes the expectation operation. The root-mean-squared (RMS) error $\varepsilon = [\langle|\hat{\theta}-\theta|^2\rangle]^{1/2}$ is also used in the following sections.

\subsection{Performance Evaluation of Proposed Method}
\begin{figure}[tb]
    \centering
    \includegraphics[width =0.7\linewidth]{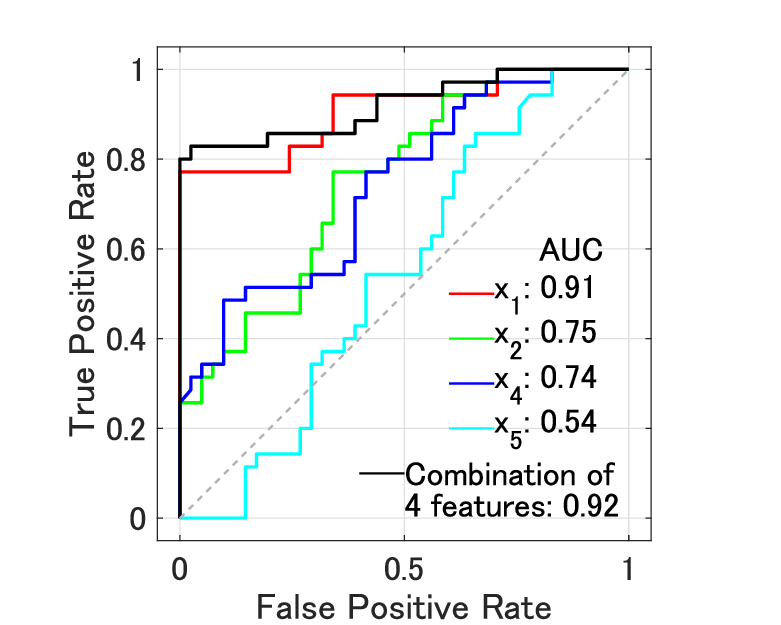}
    \caption{Characteristic analysis for the receiver operation of the binary classification.}
    \label{Fig:ROC_result.eps}
\end{figure}

The mean $\pm$ standard deviation and p-values of the five features are listed in Table~\ref{tbl:Characteristics of the five features taken from dataset of waveforms}, where the significance level was set to $p < 0.05$. All features except $x_3$ were confirmed to satisfy the significance level. The ROC curves calculated using different combinations of the features satisfying $p < 0.05$ are displayed in Fig.~\ref{Fig:ROC_result.eps}, for which a 5-fold cross-validation was adopted. To compare the performance of the ROC curves with different features, the area under the curve (AUC) was used. Feature $x_1=|D(f_0)|$ is seen to be a good indicator and its AUC is almost the same as that of a combination of the four features $x_1$, $x_2$, $x_4$, and $x_5$. For this reason, we used only $x_1$ without any other features for the regression model for the proposed binary classification; the adopted regression coefficients were $\beta_0=-5.60$, $\beta_1=6.93$ $\mathrm{mm}^{-1}$, and $\beta_2=\beta_3=\beta_4=\beta_5=0$. For the entire dataset obtained from the five participants, the confusion matrix of the logistic regression model is presented in Table~\ref{tbl:Performance of logistic}, and the average accuracy in the first step was $82.0\%$.

\begin{table}[tb]
\caption{Characteristics of the five features.}
\centering
\begin{tabular}{c|rrr}
\hline
    Feature&Class 1 & Class 2 &p-value\\
    \hline \hline
    $x_1$ (mm)  & $1.09$ $\pm$ $0.31$ & $0.54$ $\pm$ $0.19$ & $< 0.05$ \\
    $x_2$  & $0.33$ $\pm$ $0.15$ & $0.51$ $\pm$ $0.25$ & $< 0.05$ \\  
    $x_3$ (rad) & $-0.57$ $\pm$ $0.97$ & $-0.55$ $\pm$ $1.18$ & $0.72$ \\
    $x_4$  & $0.18$ $\pm$ $0.08$ & $0.32$ $\pm$ $0.16$ & $< 0.05$ \\
    $x_5$ (rad) & $-0.22$ $\pm$ $1.02$ & $0.13$ $\pm$ $1.55$ & $< 0.05$ \\
    \hline
  \end{tabular}
  \label{tbl:Characteristics of the five features taken from dataset of waveforms}
\end{table}

\begin{table}[tb]
\caption{Confusion matrix of the logistic regression model}
\centering
\begin{tabular}{l|l|c|c|c}
\multicolumn{2}{c}{}&\multicolumn{2}{c}{Actual}&\\
\cline{3-4}
\multicolumn{2}{c|}{}&Class 1&Class 2&\multicolumn{1}{c}{Total}\\
\cline{2-4}
\multirow{2}{*}{Estimated}&Class 1 & $78\%$ & $14\%$ & $92\%$\\
\cline{2-4}
&Class 2& $22\%$ & $86\%$ & $108\%$\\
\cline{2-4}
\end{tabular}
\label{tbl:Performance of logistic}
\end{table}

We discuss next the second step of the two regression models. To evaluate the effectiveness of the proposed method using all features of the respiratory harmonics (not only $x_1$), were analyzed the accuracy for each of four methods, specifically, (a) a one-step method using only $x_1$ without the hierarchical model, (b) a two-step method using only $x_1$ with the hierarchical model, (c) a one-step method using $x_1$, $x_2$, $x_4$, and $x_5$ without the hierarchical model, and (d) the proposed two-step method using $x_1$, $x_2$, $x_4$, and $x_5$ with the hierarchical model.

\begin{figure}[tb]
    \begin{tabular}{cc}
      \begin{minipage}[t]{0.45\hsize}
        \centering
        \includegraphics[keepaspectratio, scale=0.35]{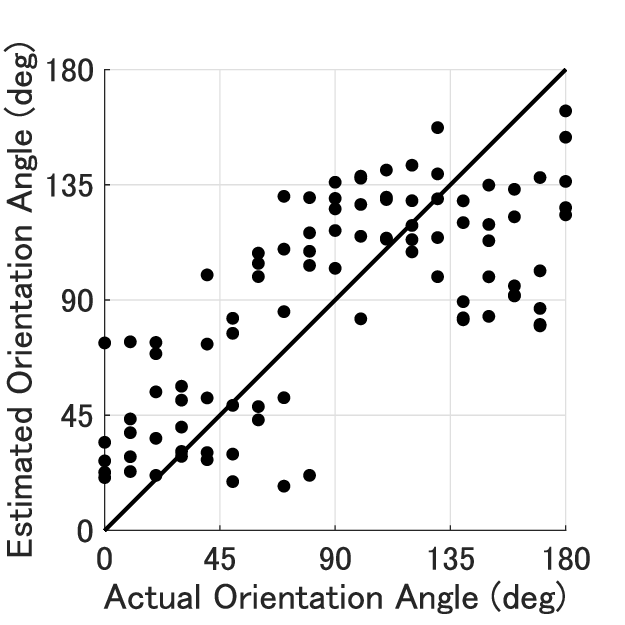}
        \subcaption{One-step method using $x_1$.}
        \label{a}
      \end{minipage} &
      \begin{minipage}[t]{0.45\hsize}
        \centering
        \includegraphics[keepaspectratio, scale=0.35]{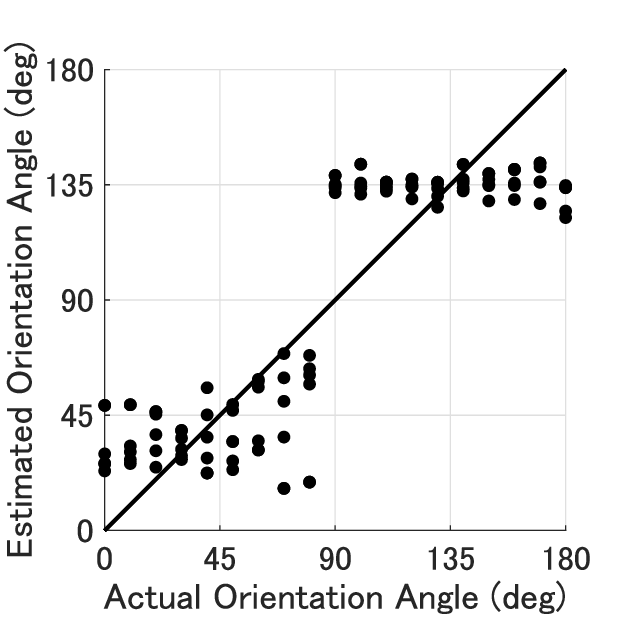}
        \subcaption{Two-step method using $x_1$.}
        \label{b}
      \end{minipage} \\

      \begin{minipage}[t]{0.45\hsize}
        \centering
        \includegraphics[keepaspectratio, scale=0.35]{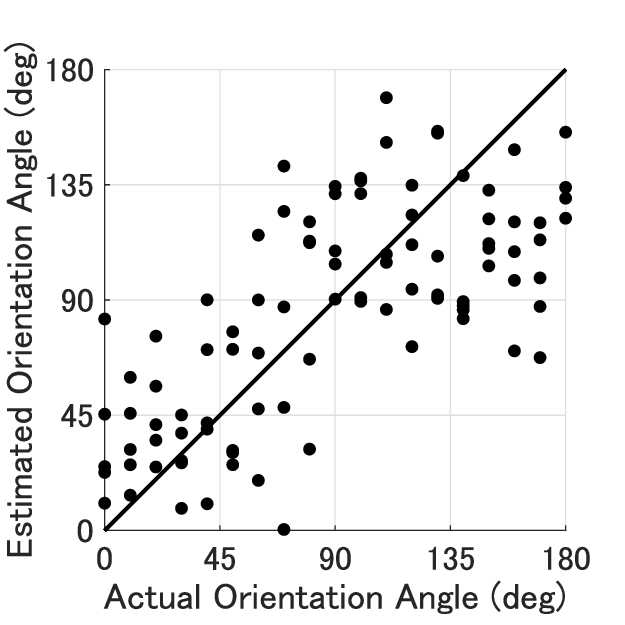}
        \subcaption{One-step method using $\vct{x}$.}
        \label{c}
      \end{minipage} &
      \begin{minipage}[t]{0.45\hsize}
        \centering
        \includegraphics[keepaspectratio, scale=0.35]{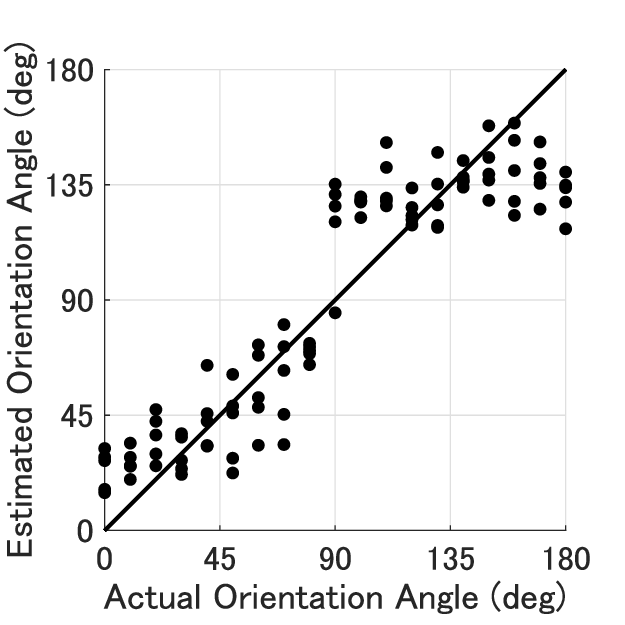}
        \subcaption{Proposed two-step method using $\vct{x}$.}
        \label{d}
      \end{minipage} 
    \end{tabular}
     \caption{Comparison of actual and estimated body orientation angles using the conventional and proposed methods.}
     \label{Fig:Conventional and proposed.eps}
\end{figure}

In the performance evaluation of these methods, a 5-fold cross-validation was also used, for which data of four out of five datasets from each participant were used to construct each model; the performance is evaluated using the data from the excluded dataset. Fig.~\ref{Fig:Conventional and proposed.eps} displays a scatter plot of $\hat{\theta}$-$\theta$ for the four methods. The CC $\rho$ and RMS error $\varepsilon$ of the methods are listed in Table~\ref{tbl:RMS error and CC}; for method (a), the values are $\rho=0.74$ and $\varepsilon=38.3^\circ$, respectively. From Fig.~\ref{Fig:Conventional and proposed.eps}(a, b), we see that method (b) performs better than (a), thus demonstrating the effectiveness of the proposed hierarchical regression model. These results indicate the validity of the binary classification in the first step of the proposed method, mainly because the component $x_1$ for front- and back-facing body orientations differ significantly, as evident in Table~\ref{tbl:Characteristics of the five features taken from dataset of waveforms}.

\begin{table}[tb]
\caption{Comparison of performances of the conventional and the proposed methods.}
\centering
\begin{tabular}{c|c|c|c}
\hline
    Methods & Features & CC $\rho$ & RMS error $\varepsilon$ (deg)\\
    \hline \hline
   (a)& $x_1$ in one step & 0.74 & 38.3 \\
   (b)& $x_1$ in two steps & 0.88 & 26.2 \\
   (c)& $x_1 \sim x_5$ in one step & 0.74 & 38.8 \\
   (d)& $x_1 \sim x_5$ in two steps & 0.91 & 23.1 \\
    \hline
  \end{tabular}
  \label{tbl:RMS error and CC}
\end{table}

The proposed method (d) improves the average estimation accuracy by factor 1.7, 1.1, and 1.7 compared with those for methods (a), (b), and (c), respectively. In Fig.~\ref{Fig:Conventional and proposed.eps}(d), the variance of the plots around the diagonal straight line $\hat{\theta}=\theta $ diminish especially for $0^\circ \leq \theta< 90^\circ$ compared with Fig.~\ref{Fig:Conventional and proposed.eps}(b). These results indicate that the features of the second and third harmonics $x_2$, $x_4$, and $x_5$ contribute in producing accurate estimations of body-orientation angle $\theta$, especially for front-facing bodies. Note that in Table \ref{tbl:RMS error and CC} the accuracy of methods (a) and (c) are almost the same, which means that harmonic features do not contribute in improving accuracy if the one-step approach is adopted.

These results indicate the importance of combining a binary classification model using $x_1$ with the regression models including respiratory harmonic features $x_2$, $x_4$, and $x_5$ in realizing accurate body orientation estimations. Compared with the conventional method, the use of the proposed method improves the coefficient of correlation $\rho$ by factor 1.2, and the average estimation accuracy $\varepsilon$ by factor 1.7. These results demonstrate the effectiveness of the proposed method in achieving a higher accuracy in estimating the body orientation by combining the fundamental component with higher-order harmonics of the respiratory motion.

\section{Conclusion}
A novel hierarchical regression model that exploits respiratory features was proposed in this study. The fundamental frequency amplitude was used in a logistic regression model to estimate whether the target person is facing towards or away from receivers. A pair of ridge regression models were then developed that combined both fundamental frequency and higher-order harmonic components. Using the proposed method, accuracies were improved by 1.7 times over those obtained by the conventional method with a simple regression model. In addition, the coefficient of correlation between the actual and estimated body-orientation angles using the conventional and proposed methods were 0.74 and 0.91, respectively. The proposed method not only contributes to accurate estimates of body orientation when using a millimeter-wave radar system, but also helps to understand the relationship between the respiratory displacement and human-body orientation.

\section*{Ethics Declarations}
This study was approved by the Ethics Committee of the Graduate School of Engineering, Kyoto University (Permit No.~202214). Informed consent was obtained from all participants in the study.

\section*{Acknowledgment}
\addcontentsline{toc}{section}{Acknowledgment}
This work was supported in part by SECOM Science and Technology Foundation, by JST under Grant JPMJMI22J2, and by JSPS KAKENHI under Grants 19H02155, 21H03427, and 23H01420. The authors thank Dr. Hirofumi Taki and Dr. Shigeaki Okumura of MaRI Co., Ltd. for their technical advice. %We thank Richard Haase, PhD, from Edanz (https://jp.edanz.com/ac) for editing a draft of this manuscript.

\end{document}